\documentclass[12pt]{article}
\usepackage{a4}
\usepackage{amssymb}
\usepackage{graphicx}
\usepackage{epsf}
\usepackage{cite}
\newcommand{\be}{\begin{equation}}
\newcommand{\ee}{\end{equation}}
\newcommand{\bea}{\begin{eqnarray}}
\newcommand{\eea}{\end{eqnarray}}

\begin{document}
\def\C{{\mathbb{C}}}
\def\R{{\mathbb{R}}}
\def\s{{\mathbb{S}}}
\def\T{{\mathbb{T}}}
\def\Z{{\mathbb{Z}}}
\def\W{{\mathbb{W}}}
\def\Bbb{\mathbb}
\def\BZ{\Bbb Z} \def\BR{\Bbb R}
\def\BW{\Bbb W} 
\def\BM{\Bbb M} 
\def\e{\mbox{e}}
\def\BC{\Bbb C} \def\BP{\Bbb P}
\def\CP{\BC\BP}
\begin{titlepage}
\title{On the Thermodynamic Geometry of BTZ Black Holes}
\author{}
\date{Tapobrata Sarkar\thanks{\noindent tapo@iitk.ac.in}, 
Gautam Sengupta, \thanks{\noindent sengupta@iitk.ac.in}, 
Bhupendra Nath Tiwari \thanks{\noindent bntiwari@iitk.ac.in}\\
\vspace{1.0cm}
Department of Physics, \\ Indian Institute of Technology,\\
Kanpur 208016, India.}
\maketitle
\abstract{

We investigate the Ruppeiner geometry of the thermodynamic state space of
a general class of BTZ black holes. It is shown that the thermodynamic 
geometry is flat for both the rotating BTZ and the BTZ Chern Simons 
black holes in the canonical ensemble. We further investigate the 
inclusion of thermal fluctuations to the canonical entropy
of the BTZ Chern Simons black holes and show that the leading 
logartithmic correction due to Carlip is reproduced. We establish 
that the inclusion of thermal fluctuations induces a non zero scalar 
curvature to the thermodynamic geometry.

}
\end{titlepage}

\section{Introduction}\label{one}
\noindent
Over the last few decades, black hole thermodynamics 
has been one of the most intense topics of research in theoretical physics
(for a comprehensive review, see \cite{wald1}). It is by 
now well known that black holes are thermodynamical systems, which 
posess a Bekenstein-Hawking (BH) entropy, and a characteristic Hawking 
temperature, related to the surface gravity on the event horizon. 
Indeed, these quantities satisfy the four laws of black hole thermodynamics. 
However an understanding the microscopic statistical origin of the black hole entropy has been
an outstanding theoretical question. Although considerable progress has been made in 
the recent past, a clear comprehension of the statistical microstates of black 
holes still remain elusive. 

It is well known that  equilibrium thermodynamic systems posess interesting 
geometric features \cite{callen}. An interesting inner product on the
equilibrium thermodynamic state space in the energy representation was provided by Weinhold 
\cite {wein} as the Hessian matrix of the internal energy with resect to the extensive thermodynamic variables. However there was no physical interpretation asociated with this metric structure. The Weinhold inner product was later formulated in the
entropy representation by Ruppeiner \cite{rupen} into a Riemannian metric in
the thermodynamic state space. The Ruppeiner geometry was however meaningful in the 
context of equlibrium thermodynamic fluctuations of the system. The curvature
scalar obtained from this geometry signified interactions and was proportional 
to the correlation volume which diverges at the critical points of phase transitions.
The {\it Ruppeiner metric} on the thermodynamic state space, is defined as the Hessian of the 
entropy with respect to the extensive variables and is given by
\begin{equation}
g_{i,j}=-\partial_i\partial_j S(U,N^a).
\end{equation}
Here $U$ denotes the internal energy of the system, and
$N^a$ are the other extensive thermodynamic variables in the entropy 
representation. Here, $i,j$ runs over all the extensive variables. 
It is to be noted that here the volume $V$ is held fixed to provide
a physical scale. 
The Ruppeiner metric is conformaly related to the Weinhold metric 
with the inverse temperature as the conformal factor.

Although, isolated asymptoticaly flat black holes do not follow 
the usual precepts of 
extensive thermodynamic systems it is possible to consider the black hole entropy
as an extensive thermodynamic quantity provided the black hole is a part of a larger system with which it is in equilibrium. From this perspective the
geometric notions of thermodynamics may be applied to investigate the nature of
the black hole entropy. In particular the investigation of the covariant thermodynamic geometry of Ruppeiner for black holes have elucidated interesting aspects of black hole phase 
transitions and moduli spaces. This was first 
explored in the context of charged black hole configurations of $N=2$ supergravity
\cite{fgk}, and since then, several authors have attempted to understand this 
connection \cite{cai1},\cite{aman1, aman2, aman3} both for supersymmetric
as well as non-supersymmetric black holes. and five dimensional rotating black rings.
The simplest black hole system for which it is possible to analyze the 
thermodynamic  geometry is the three dimensional rotating BTZ black hole. 
Here, one may construct the Ruppeiner metric in terms of the black hole mass 
and its angular momentum (the non-rotating BTZ black hole is trivial), 
and it turns out that this metric is flat \footnote {Ricci flatness in two dimensions
implies a flat space.}
\cite{aman3}. 
\footnote{See Ref. \cite{aman2} and \cite{aman3} for a classification 
of the nature of the Ruppeiner metrics for black holes in various 
dimensions.} 

Recently Kraus and Larsen \cite{krauslarsen} and Solodukhin \cite{solo} have
shown how various properties of BTZ black holes are affected by the
addition of the gravitational Chern-Simons term to the three dimensional
Einstein-Hilbert action. In particular, they show that the black hole entropy is 
modified by the presence of this term and obtain an explicit expression for 
this modified entropy. In this context, it is imperative to re-examine the Ruppeiner geometry of the BTZ black hole, in the presence of the Chern-Simons term.  

Although the statistical origin of black hole entropy is still elusive it stands to reason that
a black hole in equilibrium with the thermal Hawking radiation at a fixed Hawking temperature
is described by a canonical ensemble. The thermodynamic geometry of the black hole entropy function has been determined with reference to the canonical ensemble. However its well known that
thermal fluctuations in the canonical ensemble generates logarithmic 
corrections to the entropy. These corrections vanish in the thermodynamic 
limit where the canonical and the microcanonical entropy are identical. 
For black holes such logarithmic corrections to the canonical entropy 
have been obtained in \cite{dasetal}. 
It is a natural question as to whether the thermodynamic
geometry of black holes are sensitive to these fluctuations. As we will show in the next few sections, thermal fluctuations indeed modify the Ruppeiner geometry of the BTZ black-holes with and
without the Chern-Simons term.

In another interesting development, Sahoo and Sen \cite{sahoosen} have
computed the BTZ black hole entropy in the presence of the Chern-Simons
and higher derivative terms \cite{senentropy}. A variant of the 
{\it attractor mechanism} involving the use of the Sen entropy function
was applied to an effective two-dimensional theory that results upon making the angular coordinate of the BTZ  solution as a compact direction for this analysis. It is indeed natural
to examine the thermodynamic geometry of BTZ Chern Simons black holes with
higher derivative terms and investigate the effect of thermal fluctuations to this
geometry. It is to be emphasized here that the thermal fluctuations in the canonical ensemble 
may be analysed through purely thermodynamic considerations. In contrast the corrections to the
black hole entropy from the $\alpha'$ corrections of higher derivative terms in the effective action
needs to be analysed through gravitational considerations. Although the structures of the corrections are simmilar they may enter with opposing signs leading to a cancellation. In addition
there should be quantum corrections following from purely quantum gravitational effects. 
We note that it is not meaningful to analyse corrections due to thermal fluctuations over and above
corrections due to quantum effects.

It is the above considerations, that we set out to explore in this paper. 
Our main result is that the thermodynamic geometry is flat for the 
rotating BTZ black hole in the presence of the Chern-Simons and 
higher derivative terms. We show that inclusion of thermal fluctuations
non-trivially modify the thermodynamic geometry of the BTZ black hole
both with and without the Chern-Simons and the higher derivative corrections.
As a by-product of our results, we show that the leading order 
correction to the canonical entropy of the BTZ black hole due to thermal fluctuations 
are reproduced in the presence of Chern-Simons terms also illustrating further
the universality of these corrections. 
 
The article is organized as follows. In section \ref{two}, we first review
some known facts about the thermodynamic geometry for BTZ black 
holes, mainly to set the notations and conventions used in this paper,
and then examine the thermodynamic geometry of BTZ black holes including
small thermodynamic fluctuations. In section \ref{three}, we examine the 
Ruppeiner geometry of the BTZ black hole in the presence of the Chern-Simons term 
\cite{solo} and show that including small fluctuations in the
analysis, the leading order correction to the entropy turns out
to be the same as that of \cite{carlip}. We then calculate the
Ruppeiner curvature scalar and verify the bound on 
the Chern-Simons coupling, as predicted by Solodukhin. Section 
\ref{four} contains some comments on higher derivative corrections
to the BTZ black hole entropy, and discussions and directions for future
investigations. Some of the calculations are unfortunately
too long to reproduce here, and whereever necessary, we have used
numerical techniques to highlight and illustrate our results. 

\section{Thermodynamic Geometry of BTZ black holes}\label{two}

In this section, we study certain aspects of the thermodynamic 
(Ruppeiner) geometry of BTZ black holes. We will use the units 
$8G_N=\hbar=c=1$ and start by reviewing the results for 
the rotating BTZ black hole and then examine the role of small
thermal fluctuations.

\subsection{Rotating BTZ black holes}

The purpose of this subsection is mainly to set the notations and
conventions that will be followed in the rest of the paper. We start
with the BTZ metric
\begin{equation}
ds^2 = -N(r)dt^2 + \frac{1}{N(r)}dr^2 + r^2\left(N^{\phi}dt + d\phi
\right)^2
\label{btz1}
\end{equation}
where $N$ and $N^{\phi}$ are the (squared) lapse and shift functions
defined by
\begin{equation}
N(r) = \frac{J^2}{4r^2}+\frac{r^2}{l^2} - M;~~~~
N^{\phi}=-\frac{J}{2r^2}
\end{equation}
with $M$ and $J$ being the mass and the angular momentum of the black
hole, and $l^2$ represents the Cosmological constant term. The BTZ 
black hole has two horizons, located at
\begin{equation}
r_{\pm}=\sqrt{\frac{1}{2}Ml^2\left(1\pm \Delta\right)}
\label{rpm}
\end{equation}
where
\begin{equation}
\Delta = \sqrt{1-\frac{J^2}{M^2l^2}}
\label{delta}
\end{equation}
The mass and angular momentum of the black hole may be expressed
in terms of $r_{\pm}$ of eq. (\ref{rpm}) as;
\begin{equation}
M = \frac{r_+^2 + r_-^2}{l^2};~~~~J=\frac{2r_+ r_-}{l}
\label{mj}
\end{equation}
The BH entropy of the ordinary BTZ black hole is given by
\begin{equation}
S=4\pi r_+ 
\label{entropy}
\end{equation}  
The Ruppeiner metric is two dimensional, and is a function of the 
black hole mass $M$ and angular momentum $J$. Explicitly, the metric
is given by
\begin{equation}
g_{ij}=-\pmatrix{\frac{\partial^2 S}{\partial J^2}&
\frac{\partial^2 S}{\partial J\partial M}\cr
\frac{\partial^2 S}{\partial J\partial M}&
\frac{\partial^2 S}{\partial M^2}}
\label{rupenmatrix}
\end{equation}
with $i,j\equiv J,M$.

We will use this general form of the Ruppeiner metric throughout this
paper. A simple calculation shows that the Christoffel symbols are
given by \footnote{Our notation is $\Gamma_{ijk}=g_{ij,k}+g_{ik,j}
-g_{jk,i}$}
\begin{eqnarray}
\Gamma_{JJJ}&=&-\frac{1}{2}\frac{\partial^3 S}{\partial J^3}~~~
\Gamma_{MMM}=-\frac{1}{2}\frac{\partial^3 S}{\partial M^3}~~~
\Gamma_{JJM}=-\frac{1}{2}\frac{\partial^3 S}{\partial M \partial J^2}
\nonumber\\
\Gamma_{JMJ}&=&-\frac{1}{2}\frac{\partial^3 S}{\partial M \partial J^2}
~~~\Gamma_{JMM}=-\frac{1}{2}\frac{\partial^3 S}{\partial J \partial M^2}
~~~\Gamma_{MMJ}=-\frac{1}{2}\frac{\partial^3 S}{\partial M^2\partial J}
\end{eqnarray}
with the symmetries relating the other components.
The only non-vanishing component of the Riemann-Christoffel curvature
tensor is $R_{JMJM}=N/D$, where
\begin{eqnarray}
N=&~&\left(\frac{\partial^2S}{\partial J^2}\right)\left[
\left(\frac{\partial^3S}{\partial M\partial J^2}\right)
\left(\frac{\partial^3S}{\partial M^3}\right) -
\left(\frac{\partial^3 S}{\partial J\partial M^2}\right)^2\right]
\nonumber\\
&+&
\left(\frac{\partial^2S}{\partial M^2}\right)\left[
\left(\frac{\partial^3S}{\partial J\partial M^2}\right)
\left(\frac{\partial^3S}{\partial J^3}\right) -
\left(\frac{\partial^3 S}{\partial M\partial J^2}\right)^2\right]
\nonumber\\
&+&
\left(\frac{\partial^2S}{\partial J\partial M}\right)\left[
\left(\frac{\partial^3S}{\partial M\partial J^2}\right)
\left(\frac{\partial^3S}{\partial J\partial M^2}\right) -
\left(\frac{\partial^3S}{\partial J^3}\right)
\left(\frac{\partial^3S}{\partial M^3}\right)\right]
\label{numerator}
\end{eqnarray}
and 
\begin{equation}
D= 4\left[ 
\left(\frac{\partial ^{2}S}{\partial {J}^{2}}\right) 
\left(\frac{\partial^{2}S}{\partial{M}^{2}}\right)-
\left(\frac {\partial ^{2}S}{\partial J\partial M}\right)^{2}
\right]
\label{denominator}
\end{equation}
The Ricci scalar is 
\begin{equation}
R=\frac{2}{{\mbox{det}}g}R_{JMJM}
\label{ricciscalar}
\end{equation}
It is easy to compute the Ricci scalar by using
\begin{equation}
r_+=\frac{1}{2}\left[\sqrt{l\left(Ml+J\right)} + 
\sqrt{l\left(Ml-J\right)}\right]
\label{ricci1}
\end{equation}
Using eq. (\ref{ricci1}) in eqs. (\ref{entropy}), (\ref{ricciscalar}), 
(\ref{numerator}) and (\ref{denominator}), it can be easily shown
that the Ricci scalar vanishes identically \cite{aman3}.  
 
We might point out here that in \cite{cai1}, from considerations
of the laws of black hole thermodynamics, the authors have
argued that the internal energy of a charged or rotating black hole 
might not always be equal to its mass. Although we are not in full
agreement with the arguments of \cite{cai1}, we have checked
nevertheless that a modification of the internal energy of the
rotating BTZ black hole in lines with \cite{cai1} does not change
the observation above.

\subsection{Inclusion of thermal fluctuations}

We will now discuss the Ruppeiner geometry of
BTZ black holes including thermal fluctuations about the equilibrium.
As is well known, any thermodynamical system, considered as a
canonical ensemble has logarithmic and polynomial corrections to
the entropy \cite{huang}. 
These considerations apply to black holes as well 
(considered as a canonical ensemble), and the specific forms of
the logarithmic and polynomial corrections has been calculated for
a wide class of black holes in \cite{dasetal}. It is to be noted that
the applicability of this analysis presupposes that the canonical ensemble is 
thermodynamicaly stable. This requires a positive specific heat or correspondingly
the Hessian of the entropy function must be negative definite.

The microcanonical entropy for any thermodynamical system, incorporating such 
corrections, is \cite{huang}
\begin{equation}
S = S_0 - \frac{1}{2}{\mbox{ln}}\left(CT^2\right)
\label{correctedbtz}
\end{equation}
where $S_0$ is the entropy calculated in the canonical ensemble, and $S$ is the corrected microcanonical entropy. $C$ is the specific heat, and it is understood that 
appropriate factors of the Boltzmann's constant are included
to make the logarithm dimensionless. The approximation is valid only in the regime where
thermal fluctuations are much larger than quantum fluctuations.
In \cite{dasetal}, 
the BTZ black hole was analysed in this framework and eq. (\ref{correctedbtz}) reproduces the
leading order correction to the entropy as 
obtained in \cite{carlip}. It is then a natural question as to how the Ruppeiner geometry for the BTZ black hole is modified due to the thermal fluctuations in the canonical ensemble and this is what we will analyse in the rest of this section. 

The Ruppeiner metric for the corrected entropy for the BTZ black hole
of eq. (\ref{correctedbtz}) can be calculated using the equations
(\ref{numerator}), (\ref{denominator}) and (\ref{ricciscalar}). 
Since the expressions involved are lengthy, we will set the cosmological 
constant $l=1$ .  The Hawking temperature of the BTZ black hole is given by
\begin{equation}
T_H=\frac{1}{2\pi}\left[\frac{r_+^2 - r_-^2}{r_+}\right]
\end{equation}
which can be readily expressed in terms of the entropy of eq. (\ref{entropy})  
as
\begin{equation}
T_H=\frac{S}{8\pi^2} - \frac{8\pi^2J^2}{S^3}
\label{hawkingbtz}
\end{equation}
The specific heat is 
\begin{equation}
C = \left(\frac{\partial M}{\partial T}\right)_J = 
\frac{S\left(S^4 - 64\pi^4J^2\right)}{S^4 + 192\pi^4J^2}
\label{spheatbtz}
\end{equation}
The specific heat is positive and this ensures that the stability of the corresponding canonical ensemble. Alternatively the Hessian 
of the internal energy ( ADM mass) with respect to the extensive variables in the energy representation is given as
$$ \mid\mid \frac {\partial^2 M}{\partial X_i \partial X_j}\mid\mid
=\frac {1}{S^2l^2}- \frac {64 \pi^4 J^2}{S^6}$$. 
This is positive provided $\frac {J}{S^2}\ < \ 1$ ensuring the 
thermodynamic stability of the corresponding BTZ black hole. 
It is to be noted that this condition also governs the situation
away from extremality. Substituting the expressions of (\ref{hawkingbtz}) and 
(\ref{spheatbtz}) in (\ref{correctedbtz}), we obtain the corrected
entropy of the BTZ black hole, and the Ruppeiner metric for 
this entropy. The expression of the curvature scalar of this 
metric is far too complicated to present here, so we present the 
results numerically.

First, we consider the Ruppeiner metric with just the leading logarithmic
correction of \cite{carlip}. In this case, the analysis is simplified
and (\ref{correctedbtz}) reduces to
\begin{equation}
S = S_0 - \frac{3}{2}{\mbox{ln}}S_0
\label{correctedbtzlog}
\end{equation}

\begin{figure}[htb]
\centering
\includegraphics[width=8cm,angle=-90]{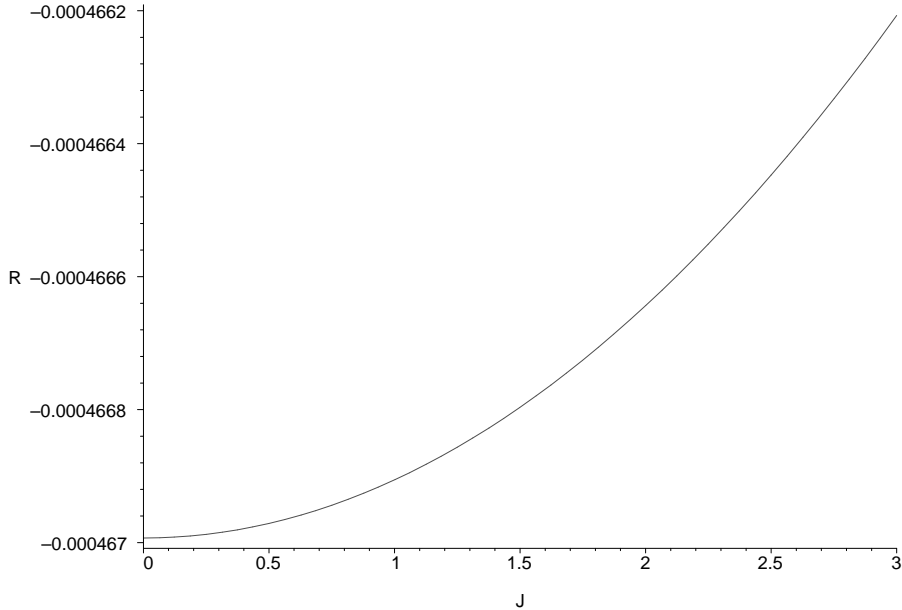}
\caption{The Ricci scalar $R$ of the Ruppeiner metric for
the BTZ black hole, as a function of the angular momentum $J$,
with only the logarithmic correction to the entropy 
(eq. \ref{correctedbtzlog}) being taken
into account. The mass $M$ has been set to $100$.}
\label{fig1}
\end{figure}

Figure (\ref{fig1}) shows the curvature scalar of the Ruppeiner
metric, $R$, plotted against the angular momentum $J$ for
$M=100$, where we have taken only the logarithmic correction 
of eq. (\ref{correctedbtzlog}) to the entropy 
into account. We have restricted to small values of $J$, so
that we are far from extremality, i.e in the regime where these results
are valid. Indeed, for near extremal BTZ black holes (i.e for very
low temperatures), our analysis is not valid \cite{dasetal}.
From fig. (\ref{fig1}), we see that in this case, the curvature scalar is
not positive definite, and indeed, by extending the values of $J$, it
is seen that the curvature scalar goes to zero at $J$ increases towards
its extremal value. However, we must point out that our calculations that
lead to this result can only be trusted when the black hole is far from
extremality. Also note that even at zero angular momentum, there is a
small but finite value of the curvature scalar. This indicates that
even at zero angular momentum, the statistical system is interacting,
once small fluctuations are included. This should be contrasted with
the non-rotating BTZ black hole which is a non-interacting system
even when small fluctuations are included. We have checked that
increasing the value of $M$, the value of the curvature scalar
becomes smaller, while preserving the shape of the graph.
 
\begin{figure}[htb]
\centering
\includegraphics[width=8cm,angle=-90]{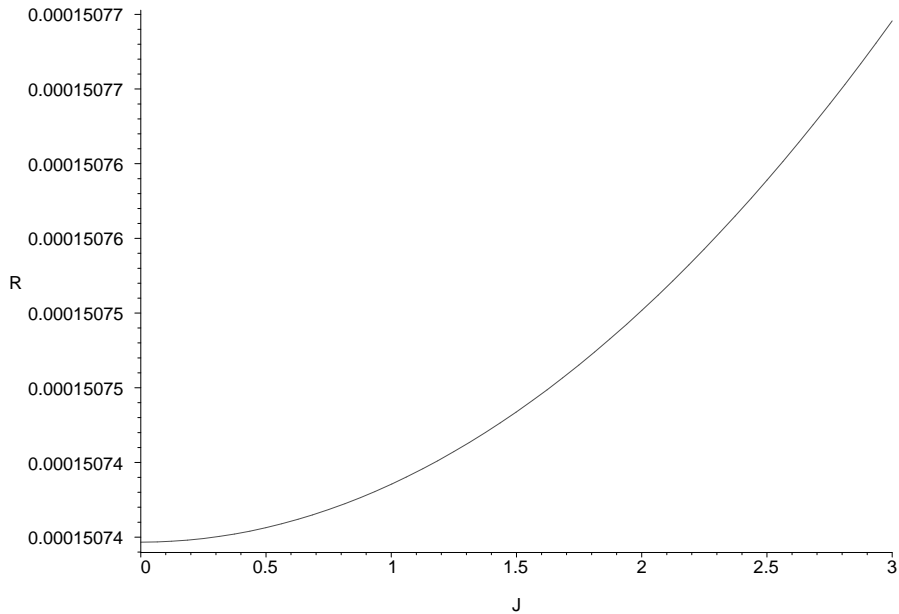}
\caption{The Ricci scalar $R$ of the Ruppeiner metric for
the BTZ black hole, as a function of the angular momentum $J$,
including small fluctuations (eq. \ref{correctedbtz}). 
The mass $M$ has been set to $100$.}
\label{fig2}
\end{figure}

Figure (\ref{fig2}) shows the Ruppeiner curvature scalar plotted
against the angular momentum, calculated using eq. (\ref{correctedbtz}).
Interestingly, in this case, the Ruppeiner scalar is 
positive definite. Again, we have restricted ourselves to values
of $J$ small compared to $M$ (i.e far from extremality) where 
our results can be trusted.
 
\section{BTZ black holes with the Chern-Simons term}\label{three}

Recently, Kraus and Larsen \cite{krauslarsen} and Solodukhin 
\cite{solo} have studied gravitational anomalies for three-dimensional
gravity in the presence of the Chern-Simons
term. Indeed, the BTZ black hole is a bonafide solution to the 
gravitational action that included both the Einstein-Hilbert and the
Chern-Simons term. We will henceforth refer to the BTZ black hole 
with the Chern-Simons term as the
BTZ-CS black hole.In \cite{krauslarsen},\cite{solo}, the entropy of BTZ-CS 
black holes have been analysed, and these authors have derived an 
expression for the entropy, which differes from the entropy of 
the ``usual'' BTZ black hole, eq. (\ref{entropy}). 
The modified entropy for the BTZ-CS black hole is
\begin{equation}
S=4\pi\left(r_+ - \frac{K}{l}r_-\right)
\label{btzcs}
\end{equation}
where $K$ is the Chern-Simons coupling. The extra term in eq. (\ref{btzcs})
as compared to eq. (\ref{entropy}) is the contribution from the 
Chern-Simons term and has very interesting properties. In particular,
\cite{solo} predicts a stability bound 
\begin{equation}
|K| \leq l
\label{stability}
\end{equation}
on the Chern-Simons coupling, from physical considerations. In view of
the above, it is natural to ask what type of Ruppeiner geometry is
seen by the BTZ black hole in the presence of the Chern-Simons term
and it is this issue that we address in this section. 

It is important to remember here that the usual mass and angular momentum
of the BTZ black hole is modified in the presence of the Chern-Simons
term. This may  be calculated by integrating the modified stress tensor
of the theory using the Fefferman-Graham expansion of the BTZ metric
and reads \cite{solo}
\begin{equation}
M = M_0 - \frac{K}{l^2}J_0; ~~~~
J= J_0 - KM_0
\label{mjcs}
\end{equation}
where $M_0$ and $J_0$ are the the mass and angular momentum of the usual
BTZ black hole of eqn. (\ref{mj}). We have calculated the Ruppeiner
metric for the BTZ black hole (with the thermodynamic coordinates
now being $M$ and $J$, rather than $M_0$ and $J_0$) in the presence 
of the Chern-Simons term, taking into account the modifications of the 
mass and angular momentum as in eq. (\ref{mjcs}). \footnote {We have set
the cosmological constant $l=1$ for simplicity.} Writing the
entropy as
\begin{equation}
S=2\pi\left[\sqrt{\left(1-K\right)\left(M+J\right)}
+\sqrt{\left(1+K\right)\left(M-J\right)}\right]
\label{btzcsentropy}
\end{equation}
it is easy to calculate the geometric quantities. The expressions
leading to the calculation of the Ricci scalar  
are not important, and we simply point out that the curvature scalar
for this geometry turns out to be zero, i.e, the Ruppeiner geometry of
the BTZ-CS black hole is flat showing that it is a non interacting statistical system. 
This is the main result of this subsection.  

\subsection{BTZ-CS black holes with small fluctuations}

We will now discuss some thermodynamic properties of the BTZ-CS black
holes, treating the system as a canonical ensemble. We allow for
small thermal fluctuations of the system considered as a canonical ensemble, and study the thermodynamic geometry of the BTZ-CS black hole in lines with our treatment of the usual
rotating BTZ black hole described earlier 

As before, we would like to analyse the Ruppeiner metric for the
BTZ-CS black hole, with the entropy now being given by eq. (\ref{btzcs}). 
Again, for ease of notation, we set the
cosmological constant $l=1$.
We begin by expressing the outer and inner horizons of the BTZ-CS
black hole as 
\begin{eqnarray}
r_+=\frac{1}{2}\left[\sqrt{M_0 + J_0} + \sqrt{M_0 - J_0}\right]
\nonumber\\
r_-=\frac{1}{2}\left[\sqrt{M_0 + J_0} - \sqrt{M_0 - J_0}\right]
\end{eqnarray}
When expressed in terms of the corrected mass and angular momentum of
eq. (\ref{mjcs}), these expressions become
\begin{eqnarray}
r_+=\frac{1}{2}\left[\sqrt{\frac{M+J}{1-K}} + \sqrt{\frac{M-J}{1+K}}
\right]\nonumber\\
r_-=\frac{1}{2}\left[\sqrt{\frac{M+J}{1-K}} - \sqrt{\frac{M-J}{1+K}}
\right]
\end{eqnarray}
The equation for the entropy, given by (\ref{btzcs}), can now be solved
to obtain the mass $M$ in terms of $S$ and $J$, and gives
\begin{equation}
M=\frac{1}{2K^2}\left[\left(2KJ+\frac{S^2}{4\pi^2}\right)+
\left[\left(2KJ+\frac{S^2}{4\pi^2}\right)^2-4K^2\left(\frac{S^4}{64\pi^4}
+\frac{S^2KJ}{4\pi^2}+J^2\right)\right]^\frac{1}{2}\right]
\end{equation}
The temperature of the BTZ-CS black hole, given by $\left(\frac
{\partial M}{\partial S}\right)_J$ may be obtained from the expression
for $M$, and is given by
\begin{equation}
T=\frac{SK^2\left[S^2\left(1-K^2\right)+8JK\pi^2\left(1-K^2\right)
+\left[S^2\left(1-K^2\right)\left(S^2+16\pi^2JK\right)\right]^{1\over 2}
\right]}{4\pi^2\left[S^2\left(1-K^2\right)\left(16\pi^2KJ+S^2\right)\right]}
\end{equation}
The specific heat may be calculated from the expression
\begin{equation}
C=\left(\frac{\partial M}{\partial T}\right)_J=
\frac{T}{\left(\frac{\partial T}{\partial S}\right)_J}
\end{equation} 
and is evaluated as
\begin{equation}
C=\frac{S\alpha\left[\beta + \left(8KJ\pi^2 + S^2\right)
\left(1-K^2\right)\right]}{\alpha\beta + S^2\left(1-K^2\right)
\left(S^2 + 24KJ\pi^2\right)}
\label{spheatbtzcs}
\end{equation}
where $\alpha = 16\pi^2KJ + S^2$ and $\beta = \left(S^2\left(1-K^2\right)\alpha\right)^{\frac{1}{2}}$. It may be checked that the secific heat is positive ensuring local thermodynamic stability.
Using eq. (\ref{spheatbtzcs}), we calculate the correction
to the canonical entropy including small thermal fluctuations of the statistical 
system and this leads to,
\begin{equation}
S=S_0 - \frac{1}{2}{\mbox{ln}}CT^2
\label{btzcsfull}
\end{equation}
where $S_0$ is the entropy (\ref{btzcs}) 
of the BTZ-CS in the canonical ensemble.
We approximate (\ref{btzcsfull}) in the limit of large entropy, 
following \cite{dasetal}. It may be easily examined that
in the limit of $S \gg J^2$ which is the stability bound, the above formula reduces to
\begin{equation}
S=S_0 - \frac{3}{2}{\mbox{ln}}S_0
\label{btzcslog}
\end{equation}
It is interesting to note that the factor of $\frac{3}{2}$, first
calculated in \cite{carlip} is reproduced for the BTZ-CS black hole
as well. Illustrating the seeming universality of this factor.
This is one of the main result of this subsection.

\begin{figure}[htb]
\centering
\includegraphics[width=8cm,angle=-90]{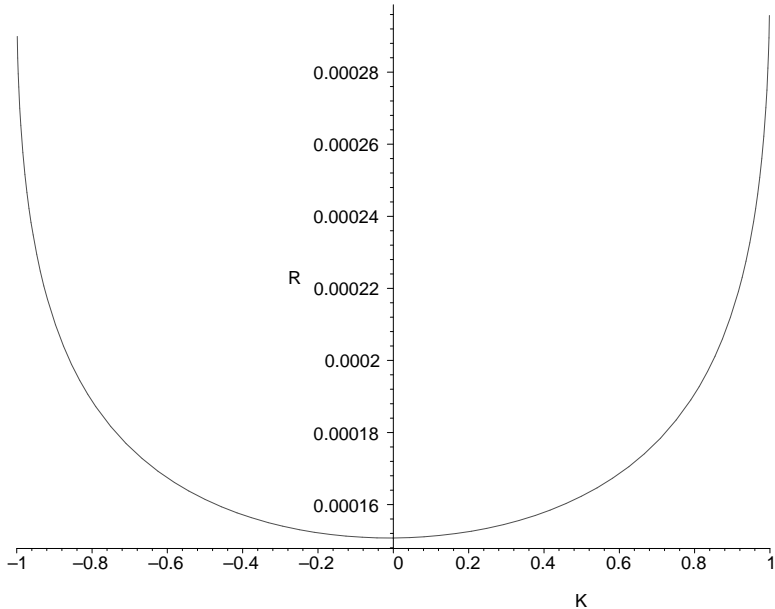}
\caption{The Ricci scalar $R$ of the Ruppeiner metric for
the BTZ-CS black hole, as a function of the Chern-Simons 
coupling $K$, with only the logarithmic correction to the entropy 
being taken into account. The mass $M$ has been set to $100$ and
we have set $J=1$ to ensure that we are far from extremality.}
\label{fig3}
\end{figure}

We now calculate the Ruppeiner geometry correspoding to the modified entropy of the 
BTZ-CS black hole with 
thermal fluctuations. As in the last section, we first present the
numerical result for the Ricci scalar using the leading order
correction of eq. (\ref{btzcslog}). 

This is depicted in fig. (\ref{fig3}). In this analysis, 
we have set $M=100$ and $J=1$, to ensure that we are far from
extremality. The Ricci scalar is positive definite in this case.  
The Ricci scalar diverges for $|K| = 1$. More appropriately, since we 
had set the cosmological constant to unity, it is not difficult to see
that the bound on $K$ from the Ruppeiner geometry is $|K| \leq l$
where $l$ is the cosmological constant. This is of course
as expected, since the entropy (\ref{btzcsentropy}) becomes
unphysical beyond this limit.

\begin{figure}[htb]
\centering
\includegraphics[width=8cm,angle=-90]{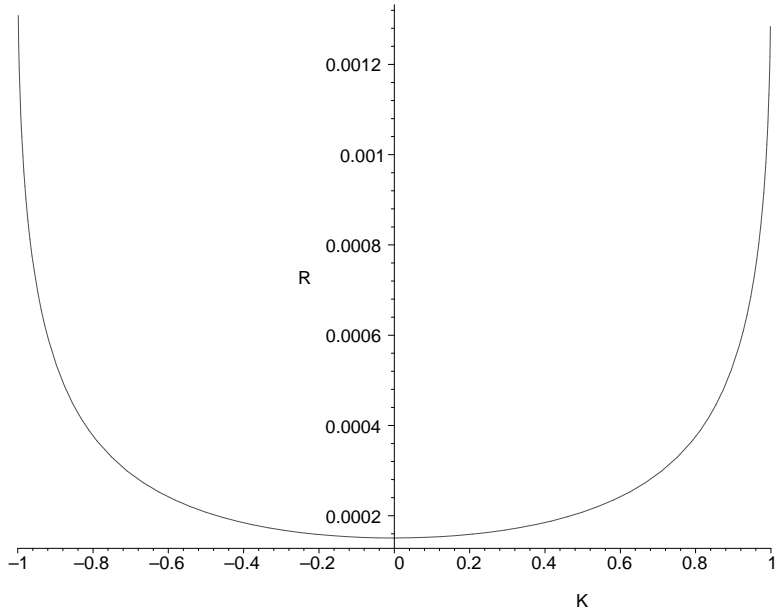}
\caption{The Ricci scalar $R$ of the Ruppeiner metric for
the BTZ-CS black hole, as a function of the Chern-Simons 
coupling $K$. The mass $M$ has been set to $100$ and
we have set $J=1$ to ensure that we are far from extremality.}
\label{fig4}
\end{figure}

In fig. (\ref{fig4}), we present the result for the Ricci 
scalar of the Ruppeiner geometry taking into account the full
correction of eq. (\ref{btzcsfull}). Again, as a function of
$K$, the curvature scalar is positive definite and the 
graph has the same qualitative features as in fig. (\ref{fig3}).
  
For the sake of completeness, we have also numerically evaluated the 
Ricci scalar of the Ruppeiner metric for the BTZ-CS black hole
as a function of the angular momentum, and studied its behaviour.
The plots in this case are qualitatively the same as in
fig. (\ref{fig1}) and fig. (\ref{fig2}) and we do not discuss 
them further.
 
\section{Discussions and Conclusions}\label{four}

In this article, we have mainly investigated the thermodynamic
geometry of a class of  BTZ black holes, both with and without the Chern-Simons
term. We have shown that the Ruppeiner geometry remains flat even with the 
introduction of the Chern-Simons term, as it was without this term. However,
introducing small thermal fluctuations in the analysis produces a non-zero
Ricci scalar for the thermodynamic geometry, and we have calculated 
this quantity for some special cases. As a byproduct of our calculations,
we have shown that the leading logarithmic correction to the 
canonical entropy of the BTZ-CS black hole retains the same form as for
the ordinary rotating BTZ black hole thus illustrating the universality of this
correction. We should mention here that the validity of this analysis depends on
the local thermodynamic stability which is ensured by a positive specific heat for the
BTZ and the BTZ-CS black holes. This is also generaly true for charged and rotating charged black holes. It would be intetesting to extend our analysis to other black holes and investigate
the subtle interplay between the corrections due to thermal fluctutaions and $\alpha"$
corrections resulting from higher derivative terms. It is expeted that the corresponding thermodynamic geomteries would be sensitive to these corrections. Furthermore thermodynamic
geometries provide a direct way to analyse critical points of black hole phase transitions which
is an area of current interest. This may have important implications for black holes
in string theory and the geomtry of moduli spaces. Some of these issues will be investigated in future.

A few comments are in order here. It is clear that our analysis will
be similar for BTZ black holes with higher derivative corrections. 
As shown in \cite{krauslarsen} and \cite{sahoosen}, the form of
the entropy for the BTZ-CS black hole remains the same in the presence
of the higher derivative corrections, and it is the central charge of
the underlying conformal field theory that is modified. Hence, we expect
qualitatively similar results for the thermodynamic geometry of BTZ-CS black holes with 
higher derivative corrections.  We have explicitly verified this.

As we have pointed out earlier, leading logarithmic correction to the black hole entropy arises
from various sources. The black hole considered as a canonical ensemble admits such corrections
to the entropy due to standard thermal fluctuations. The effect of such fluctuations
may be analysed from purely thermodynamic considerations. It is to be understood that
these fluctuations vanish in the thermodynamic limit of large systems where the canonical and the
microcanonical entropy becomes identical. Apart from these the black hole entropy also admits logarithmic corrections due to presence of higher derivative terms to the gravitational action from
the perspective of low energy effective field theories resulting from some underlying theory of
quantum gravity. These higher derivative corrections are accessible to analysis through purely
gravtitational considerations like Walds formula or through gravitational anomalies. It is a meaningful exercise to analyse these two corrections simultaneously and in certain cases leads to
a cancellation. However corrections due to purely quantum effects must be considered separately.
Lacking a viable fundmaental theory of quantum gravity these quantum corrections still need to be
elucidated.

We should also remark here that as pointed out in \cite {solo} that the modification of the entropy due to the gravitational Chern-Simons term being dependent on the radius of the inner horizon
seems to probe the black hole interior. This is in contarst to the higher derivative $\alpha'$ corrections which are only dependent on the radii of the outer horizon. This seems to indicate that
contrary to the existing point of view certain degrees of freedom may be associated with the black hole interior. This may have implications for space time holography and is an important
issue for future investigations.

\newpage
\section{Acknowledgements}\label{six}

We are grateful to Ron-Gen Cai for helpful email correspondence. 
All of us would like to thank Geetanjali Sarkar for computational 
support and V. Subrahmanyam for discussions. GS would like to thank 
Sutapa Mukherji for a reference. BNT acknowldges CSIR, India, 
for financial support under the research grant 
CSIR-SRF-9/92(343)/2004-EMR-I.

\end{document}